# First non-icosahedral boron allotrope synthesized at high pressure and high temperature

Irina Chuvashova[a,b*], Elena Bykova[b], Maxim Bykov[b], Vitali Prakapenka[c], Konstantin Glazyrin[d], Mohamed Mezouar[e], Leonid Dubrovinsky[b], Natalia Dubrovinskaia[a]

[a] *Material Physics and Technology at Extreme Conditions, Laboratory of Crystallography, University of Bayreuth, D-95440 Bayreuth, Germany*
[b] *Bayerisches Geoinstitut, University of Bayreuth, D-95440 Bayreuth, Germany*
[c] *Center for Advanced Radiation Sources, University of Chicago, 9700 South Cass Avenue, Argonne, IL 60437, USA*
[d] *Photon Science, Deutsches Elektronen-Synchrotron, Notkestrasse 85, D-22607 Hamburg, Germany*
[e] *European Synchrotron Radiation Facility, BP 220 F-38043 Grenoble Cedex, France*

*\*Correspondence to:* Natalia.Dubrovinskaia@uni-bayreuth.de, *irina.chuvashova@gmail.com*

**Abstract**

Theoretical predictions of pressure-induced phase transformations often become long-standing enigmas because of limitations of contemporary available experimental possibilities. Hitherto the existence of a non-icosahedral boron allotrope has been one of them. Here we report on the first non-icosahedral boron allotrope, which we denoted as ζ-B, with the orthorhombic α-Ga-type structure (space group *Cmce*) synthesized in a diamond anvil cell at extreme high-pressure high-temperature conditions (115 GPa and 2100 K). The structure of ζ-B was solved using single-crystal synchrotron X-ray diffraction and its compressional behavior was studied in the range of very high pressures (115 GPa to 135 GPa). Experimental validation of theoretical predictions reveals the degree of our up-to-date comprehension of condensed matter and promotes further development of the solid state physics and chemistry.

# I. INTRODUCTION

Boron has been widely studied due to its complex polymorphism (see the review article [1] and references therein). All five of the hitherto experimentally established boron allotropes (α-B, β-B, γ-B, δ-B, and ε-B) [1-6] belong to the family of icosahedral solids: there structures are based on various arrangements of $B_{12}$ icosahedra, since three valence electrons of boron atoms are insufficient to form a simple covalent structure. However, van Schnering and Nesper [7] suggested the α-Ga structure to be an alternative to the boron structure. According to these authors, the α-Ga structure built up from open polyhedral fragments is strongly related to the closed polyhedral boron clusters $B_n$. Their extended Hückel calculations supported the α-Ga model and showed that the α-Ga structure is an appropriate covalent three-electron arrangement and is not electron-deficient [7].

The pressure-temperature (PT) experimental phase diagram of boron is currently limited to 20 GPa and 3500 K [6,8-13], but there are a number of theoretical predictions concerning the behavior of boron at higher pressures and temperatures [14-21]. Häusserman et al. (2003) [15] predicted the phase transition from α-B to α-Ga-structured phase, accompanied by a nonmetal-metal transition at 74 GPa. The authors [15] used *ab initio* calculations employing pseudopotentials and a plane wave basis set in the framework of the density functional theory (DFT). The α-Ga-structured boron phase was suggested to be stable up to 790 GPa; beyond this pressure a transition to the *fcc* structure had to take place according to [15]. In parallel, Segal and Arias (2003) [22] performed calculations using a method based on perturbation theory and all-electron calculations with plane-wave-basis in DFT. The α-Ga-structured boron was shown to be favorable in energy among other boron phases in the interval between 71 GPa and 500 GPa. Note that at that time γ-B was not discovered yet. Later, a comparison of the stability of γ-B with respect to the α-Ga-type boron phase led to consequent shift of their phase boundary from 78 GPa, as calculated in [14], to 89 GPa [20], and 93 GPa [19]. Calculations predict α-Ga-structured boron to be an incompressible ($K_{300}$ = 263 GPa) [18], superconducting material with strong anisotropy [23] due to its layered crystal structure [22,23]. Electric resistivity measurements [12] showed that compression of β-B leads to metallization of the material under investigation at pressures above 160 GPa and ambient temperature, and to formation of a superconducting phase above 175 GPa and 6 K [12]. The structure of metallic superconducting boron is unknown.

Although the idea of possible existence of boron with the α-Ga structure is already more than 25 years old [7], it has remained difficult to prove. First, very high pressures are required for its synthesis, as predicted by [15]; second, boron is a weak X-ray scatterer that means that HPHT experiments are not straight forward and have to be done on a synchrotron; and third, a precursor material has to be of a very high purity, what is not always easy to reach with the highly reactive boron.

The logic of our experimental approach, aimed at overcoming the challenges listed above, is as follows. As a precursor material, we used single crystals of β-B from the same synthesis batch as those which were fully characterized in our previous work [6]. The crystals were proven to be of high quality and high purity. To assure that we can follow the known β-B to γ-B transition at moderate pressures [4], we conducted first a few synchrotron XRD experiments below one Mbar to see if our results are in accordance with our own previous observations and available literature data. Further experiments at higher pressures, up to above one Mbar, aimed at investigating the HP behavior of β-B to track possible phase transitions at room temperature (Eremets et al. [12] reported a visible step in the resistance of β-B at 110 GPa) and upon heating.

Here, we report a new boron allotrope (ζ-B) with the α-Ga-type structure synthesized from β-B at pressures over 115 GPa and temperatures over 2100 K using a laser heated diamond anvil cell (DAC). Its crystal structure was determined based on single-crystal synchrotron X-ray diffraction (XRD) data. The behavior of ζ-B under compression from 115 GPa to 135 GPa was characterized using powder synchrotron XRD.

## II. EXPERIMENTAL

**Synthesis of a precursor material.** Single crystals of β-B studied in the present work at high pressures and high temperatures were synthesized using the high-pressure high-temperature technique described in detail in [6]. Their structure and purity were carefully characterized to assure the reliability of the obtained experimental results. The presence of impurities could be excluded.

**Diamond-anvil cell experiments.** The BX90-type diamond anvil cells (DAC) [24] made at Bayerisches Geoinstitut (Bayreuth, Germany) and Boehler-Almax type [25] beveled diamonds with the culet diameters of 120 µm were used in high pressure experiments. Rhenium gaskets were squeezed between the anvils to make an indentation with the thickness of about 20 µm. Then in the center of the indentations, round holes of about 60 µm in diameter were drilled. Two β-B crystals were placed into these chambers. Sizes of the crystals were about $10 \times 10 \times 15$ µm$^3$ and orientation of the crystals was not specified. Neon was used as a pressure transmitting medium (PTM) and as pressure standard [26].

**Single crystal synchrotron X-ray diffraction.** Single crystals of β-B in a DAC were studied on ID27 at the European Synchrotron Radiation Facility (ESRF) and on P02.2 at PETRA III, DESY [27].

At ID27 diffraction data were collected at 293 K using the Perkin Elmer XRD1621 flat panel detector. The monochromatic radiation had the wavelength of 0.37380 Å and the crystal-to-detector distance was 383 mm. Pressure in the cells was increased stepwise up to about 115 GPa, and single-crystal diffraction data for β-B were collected at each pressure point. 160 frames in the omega scanning range of −40° to +40° (in 0.5° steps) were recorded with an exposure time of 2 s. A portable double-sided laser heating system [28] was used to heat β-B crystals under pressure in experiments at ID27 (ESRF). Upon heating entire crystals were located in the laser beam and there were no measurable temperature gradients within the samples. The temperature variation during heating did not exceed ±100 K.

At P02.2 diffraction data were collected at 293 K using the Perkin Elmer XRD1621 detector. The monochromatic radiation had the wavelength of 0.29464 Å and the crystal-to-detector distance was 439 mm. Data were collected at one pressure point at about 115 GPa on the sample laser heated at ID27. 152 frames in the omega scanning range of −38° to +38° were collected (in 0.5° steps) with an exposure time of 10 s per frame.

Integration of the reflection intensities and absorption corrections were performed using CrysAlisPro software [29,30]. The structure of γ-B was refined in the anisotropic approximation for all atoms by full matrix least-squares Due to small amount of the observed data the structure of ζ-B was refined in isotropic approximation. Refinements of the crystal structures were performed using SHELXL software [31] implemented in the WinGX software package [30]. The crystallographic data of ζ-B studied at 115 GPa have been deposited in the Inorganic Crystal Structure Database [39]. The data may be obtained free of charge from Fachinformationszentrum Karlsruhe, 76344 Eggenstein-Leopoldshafen, Germany (Fax: +49 7247 808 666; e-mail: crysdata@fiz-karlsruhe.de, http://www.fiz-karlsruhe.de/request_for_deposited_data.html) on quoting following CSD deposition number: 432572.

**Powder XRD measurements.** Samples of ζ-B were studied at room temperature in angle-dispersive mode with a wavelength of 0.2952 Å at the 13-IDD beamline at APS, Argonne. Pressure in the cells was increased from 115 to 135 GPa with a step of about 2 GPa. Diffraction images were collected at each pressure using a MAR CCD detector in the omega scanning range of −20° to +20° with an exposure time of 40 s. The images were integrated using the DIOPTAS software [32] and the resulting diffraction patterns were processed using biased model in the EXPGUI software [33] in the GSAS Software package [34].

## III. RESULTS AND DISCUSSION

In two independent DAC experiments at ID27 at ESRF single crystals of β-B were compressed up to 38(1) and 50(1) GPa, respectively. At these pressures the observed diffraction patterns perfectly match that of β-B (details of the HP structural studies of β-B will be published elsewhere). The lattice parameters and the molar volume of β-B corresponding to these two pressures are in agreement with the literature data of Sanz et al. [13], who measured the lattice parameters of β-B up to 100 GPa. Figure 1 shows the pressure dependence of the relative unit cell volume of β-B, as experimentally determined in [13] (black curve), and our two experimental points (black squares), which appear very close to the curve. After compression and the XRD measurements, the both crystals of β-B were double-side laser heated to about 2000 K. After the heating the pressure in the both DACs (determined using the equation of state of neon) increased and became 42(1) GPa in the first DAC and 58(1) GPa in the second one. In accordance with the experimental PT phase diagram of boron [6], a transition of β-B to γ-B took place, and diffraction spots of γ-B were clearly observed in the diffraction pattern taken at room temperature (RT) after heating at 42 and 58 GPa. The quality of the HP single-crystal X-ray diffraction data was sufficient to refine both the lattice parameters and atomic coordinates of γ-B (details of the refinement of the structure of γ-B at 58(1) GPa are presented in Supplementary Table I). The structure of γ-B is orthorhombic (space group *Pnnm*) and built of covalently bonded $B_{12}$ icosahedra and $B_2$ dumbbells [35].

To compare the compressional behavior of various boron allotropes and to plot the P-V data on the same graph (Figure 1), the relative unit cell volumes for all considered allotropes were normalized to the unit cell volume of β-B per atom (320 atoms in β-B were accepted following [13]). The blue curve in Figure 1 corresponds to the experimental data for γ-B from [4].Two experimental points obtained in the present study are shown by blue diamonds and lie on this curve.

In order to study the behavior of boron in a megabar pressure range, on ID27 at ESRF single crystals of β-B were first compressed up to 100(2) GPa. At this pressure at room temperature, all observed reflections were attributed to β-B. This point (the black square in Figure 1 corresponding to 100 GPa) fits well to the curve of Sanz et al. [13] extrapolated to 115 GPa (dashed black curve). After double-sided laser-heating up to 2100(100) K using portable laser heating set up, pressure increased to 115(2) GPa. The material in the heated spot changed its color from dark-reddish to black (not reflecting). The X-ray diffraction pattern changed dramatically and had to be treated as of a powder sample. Apart from reflections of Re (gasket material) and Ne (used as PTM), several new relatively weak but clearly detectable reflections were observed. Their d-spacings perfectly matched those expected for α-Ga-type structured boron, as predicted by Häusserman et al. [15]. The following orthorhombic lattice parameters were obtained at 115 GPa: $a = 2.7159(11)$ Å, $b = 4.8399(2)$ Å, $c = 2.9565(6)$ Å.

The DAC with the sample described above was transported to the P02.2 beamline at PETRA III, where combination of a small size of the X-ray beam and its short wavelength increases chances for accumulating diffraction data suitable for single-crystal structural analysis. Detailed inspection of the diffraction pattern obtained on this beamline from the heated spot at 115 GPa (Supplementary Figure 1) revealed individual single crystal reflections related to the high-pressure high-temperature boron phase. Its structure was solved; some experimental details and crystallographic data are given in Supplementary Table 1. The structure belongs to the α-Ga structure type; it has the *Cmce* space group and 8 atoms per unit cell. The unit cell parameters determined from the single-crystal data ($a = 2.7039(10)$ Å, $b = 4.8703(32)$ Å, $c = 2.9697(6)$ Å) are slightly different from those obtained from powder XRD at ID27, as they were determined using different sets of reflections, as well as different procedures of finding peaks positions, weight schemes in least square optimizations, etc. The new high-pressure boron allotrope with the α-Ga-type structure was denoted as ζ-B, sequentially after the fifth hitherto established boron allotrope, ε-B [5].

The structure of ζ-B at 115(2) GPa is presented in Figure 2. It may be described as a stacking along the (010) direction of distorted and corrugated hexagonal nets (Figure 2A) with the $3^6$ topology, in accordance with the descriptions of van Schnering et al. [7,36] and Häusserman et al. [7,36]. Within each net each B atom connects to six neighbors, and the B-B bond lengths are 1.66(1) Å, 1.72(1) Å, and 1.75(1) Å (notated in Figures 2). The bonds between the nets bring the seventh neighbor to the coordination sphere of each boron atom (Figure 2B). These B-B bonds appear to be the shortest (1.59(1) Å) in the structure of ζ-B. Thus, despite seemingly 'layered' appearance, the ζ-B structure is in fact a 3D network. Puckered fragments of the nets can be considered as open polyhedral fragments related to the closed icosahedral boron clusters.

As mentioned above, in ζ-B each boron atom has a coordination number (CN) equal to 7. In α-B a half of boron atoms have CN=6, and a half CN=7 [37], and in γ-B out of 28 atoms in the unit cell eight atoms have CN=6, and other twenty atoms have CN=7 [38]. For a complex and still controversial structure of β-B a simple count is not possible, but anyhow most of the boron atoms in this phase have CN=6. Thus, comparing the structures of the boron allotropes, we observe a tendency to a rise of the CN of boron for high-pressure polymorphs. This agrees with the empiric rule that upon pressure-induced phase transition coordination number increases [39]. As expected [40], an average interatomic distance in the first coordination sphere is the longest for ζ-B (1.68 Å) in comparison with corresponding values for α-B (1.59 Å) [37], and γ-B (1.66 Å) [38] at 115 GPa.

A phase transition of β-B to ζ-B manifested in a drastic reduction of the molar volume: at 115 GPa the molar volume of ζ-B is by ~7.5 % less than that of β-B. According to theoretical calculations [14,18,20] and our previous [6] and current experimental observations (up to about 60 GPa), γ-B is more stable at high pressures than β-B. If we compare the difference in the molar volume between γ-B and ζ-B at 115 GPa, it is only about 3.1% (see Figure 1, where the pressure dependence of the normalized volume of γ-B taken from [4] was plotted for the comparison).

It is worth mentioning here that we pressurized single crystals of β-B beyond the pressure (110 GPa), at which Eremets et al. [12] observed a kink in the room-temperature $R(P)$ (resistance *vs* pressure) curve. The authors [12] suggested the possibility that the transition of β-B to the metallic state occurs at 130 GPa. Häusserman et al. [15] proposed the α-Ga structure as a structural model for a metallic high-pressure modification of B after a phase transition of either semiconducting icosahedral α-B or β-B. We did not observe any transformations in β-B up to 115 GPa, and the transition to the α-Ga-structured phase required heating to very high temperatures. Thus, it is very improbable that ζ-B with the α-Ga structure may be associated with the metallic high-pressure modification of boron discussed in [12,15].

Further compression of the material synthesized at 115(2) GPa leads to decrease of the quality of single-crystal reflections. For this reason, the diffraction data of ζ-B obtained up to 135 GPa on 13-IDD at the APS were integrated to 1D ´2-theta´ scans. An example of a diffraction pattern at 121(2) GPa is presented in Figure 3. The unit cell parameters decreased with pressure. The linear compressibility along the *a* and *c* axes was found to be ~7(3)*$10^{-4}$ $GPa^{-1}$, and along the *b* axis - a bit lower, ~4(3)*$10^{-4}$ $GPa^{-1}$ that can be explained by the shortest B-B bonds in the (010) direction. The PV data set of ζ-B in the pressure range of 115 to 135 GPa was fitted using the second-order Birch-Murnaghan (2BM) equation of state (EoS) and gave the following EoS parameters: $V_{115}$ = 39.20(8) Å$^3$ and $K_{115}$ = 575(65) GPa; K´= 4 (fixed); $V_{115}$ is the unit cell volume and $K_{115}$ is the bulk modulus at 115 GPa and room temperature; K´ is the pressure derivative of the bulk modulus) (Table I). Use of the Vinet EoS led to the following parameters: $V_{115}$ = 39.19(8) Å$^3$ and $K_{115}$ = 577(65) GPa; K´= 4 (fixed).

The bulk moduli of α-Ga-structured boron at ambient conditions were calculated in [18,19,23]. To compare these predictions with our experimental results, we computed the theoretical values of the bulk moduli to be at 115 GPa according to these three papers (Table I). The EoSFit-7c software [41] was used. As seen, experimental values of the bulk modulus are lower than the theoretically predicted ones. Supplementary Figure 2 shows the pressure dependence of the unit cell volume of ζ-B, as experimentally determined in the present work and theoretically predicted [18,19,23]. There is a significant difference (about 4%) between the earlier [23] and recent [18,19] theoretical predictions. Our experimental results are in reasonable agreement with data of [18,19] (Supplementary Figure 2).

## IV. CONCLUSIONS

To verify theoretical predictions regarding the existence of α-Ga-structured boron and its behavior at high pressures [7,14,15,19,20], we have conducted a series of high-pressure high-temperature experiments. We demonstrated that the predicted boron allotrope [7,15] can be obtained by laser-heating of single crystals of β-B to over 2100 K at pressures above 115 GPa. This phase, which we call ζ-B, has the α-Ga-type orthorhombic structure as revealed by single-crystal X-ray diffraction. Measured precisely interatomic distances and linear compressibilities along the major crystallographic directions do not allow interpreting the structure as layered, as earlier proposed [23]. In the studied pressure range (from 115 to 132 GPa) ζ-B is less compressible than any other boron allotropes known so far. Based on our experimental data we do not see a relation between ζ-B and the metallic high-pressure modification of B discussed by [12,15].

## ACKNOWLEDGEMENTS

We thank G. Parakhonskiy for providing us with samples of β-B used here. N.D. thanks the German Research Foundation (Deutsche Forschungsgemeinschaft (DFG)) and the Federal Ministry of Education and Research (BMBF; Germany) for financial support through the DFG Heisenberg Programme (projects no. DU 954-6/1 and DU 954-6/2) and project no. DU 954-8/1 and the BMBF grant no. 5K13WC3. L.D. thanks the DFG and the BMBF (Germany) for financial support.

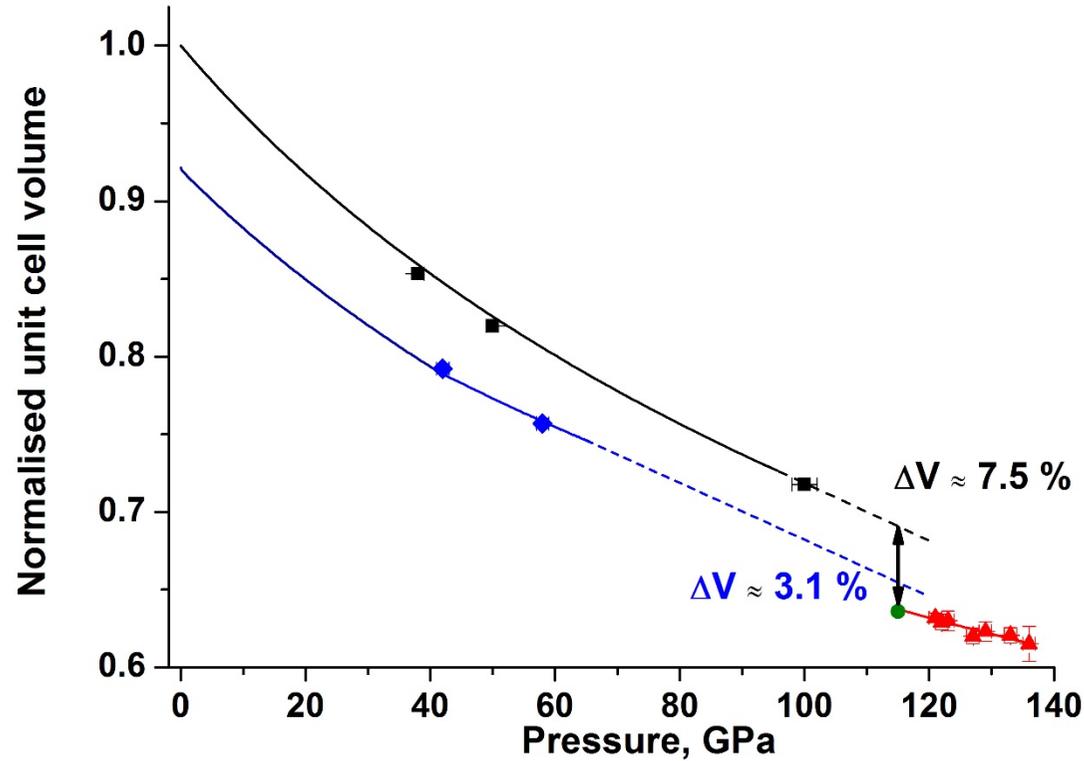

**FIG. 1. (Color online) The pressure dependence of the normalized relative unit cell volumes of three boron allotropes, β-B, γ-B and ζ-B.** The volumes were normalized on the volume of β-B per atom calculated from the experimental data of Sanz et al. [13]. The solid black curve corresponds to the Vinet equation of state of β-B up to 100 GPa from [13] ($V_0$ = 2460 Å$^3$, $K_{300}$ = 210(6) GPa, $K´$ = 2.23). Dashed black curve is its extrapolation to 120 GPa. Black squares represent our experimental data points for β-B (see text). The green circle represents the volume data for ζ-B obtained from our single-crystal XRD at 115 GPa (after laser heating of β-B at this pressure a phase transition occurred, accompanied with the volume reduction by ca.7.5 %). The red triangles correspond to the PV data of ζ-B obtained from powder XRD. Their fit with the 2BM EoS is shown by the red solid curve. The blue curve corresponds to the 3BM EoS of γ-B according to [38] (below 40 GPa: $V_0$=198.1(3) Å$^3$, $K_{300}$=227(3) GPa, K'=2.5(2)); above 45 GPa: $V_0$=192.6(3) Å$^3$, $K_{300}$=281(6) GPa, K'=2.8(9)). Its extrapolation to 120 GPa is shown by a dashed blue curve. Blue diamonds show our two experimental points (see text). The volume difference between γ-B and ζ-B at 115 GPa is ca. 3.1 %.

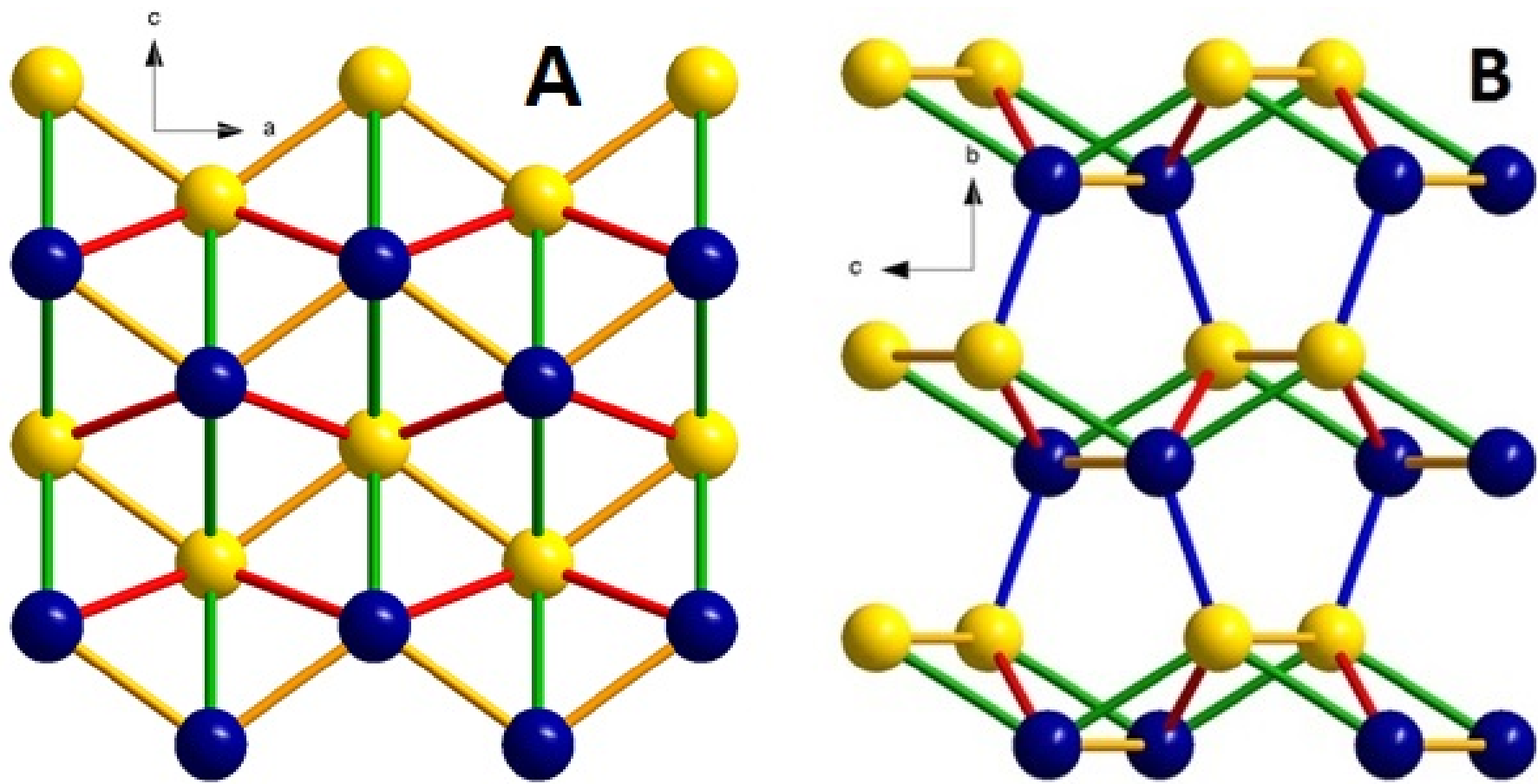

**FIG. 2. (Color online) The structure of ζ –B. A.** The projection of a fragment of one distorted and corrugated hexagonal net on the *ac* plane. Such nets are stacked along the (010) direction. Blue and yellow atoms do not lie in the same plain (blue atoms are lower and yellow ones are upper if seen along the (010) direction). Bonds with different lengths are shown in different colors: 1.66(1) Å (orange), 1.72(1) Å (red), and 1.75(1) Å (green). **B.** The projection of three nets on the *bc* plain. The lengths of bonds connecting the layers (blue) all are 1.59(1) Å.
10

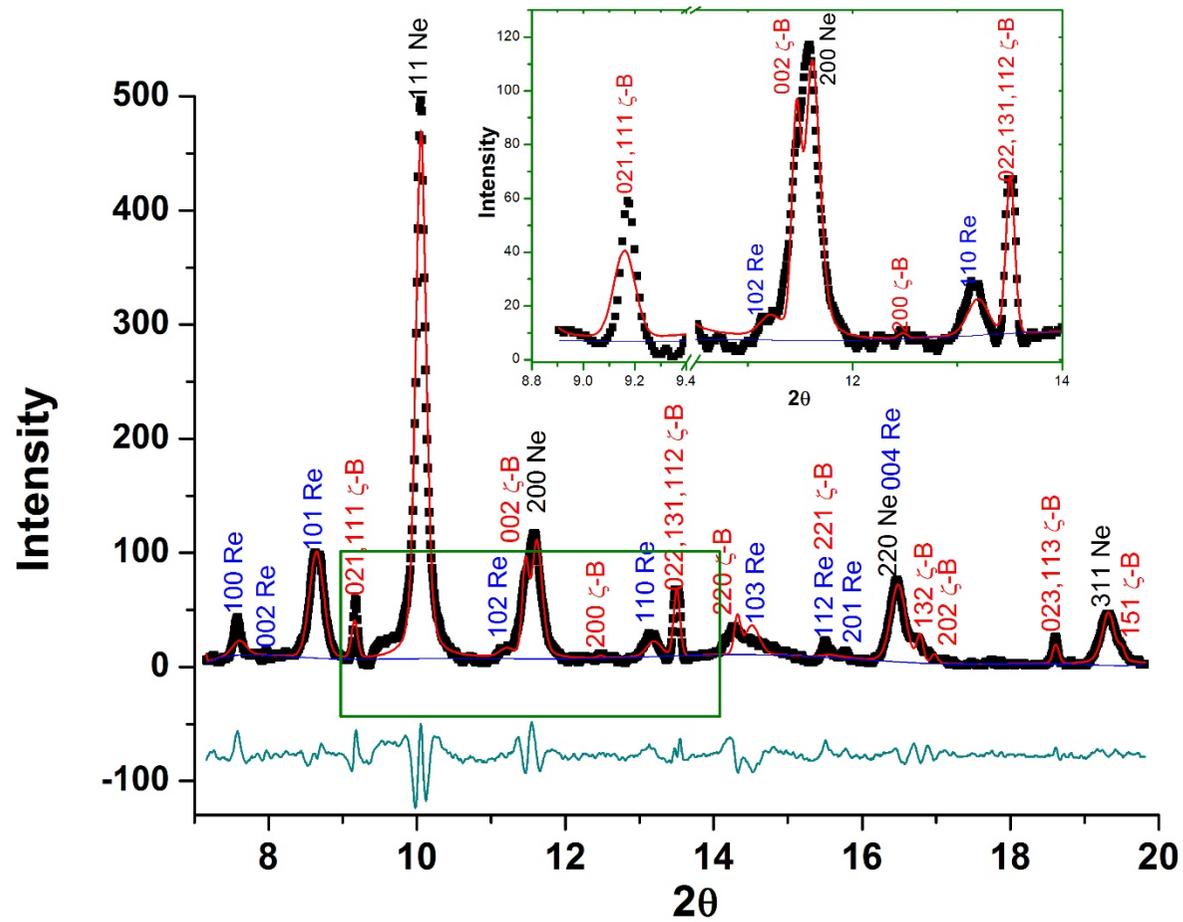

**FIG. 3. (Color online) X-ray diffraction pattern of ζ-B obtained at 121(2) GPa.** Black dots represent experimental data. Red solid line derives the refinement using biased model, blue solid line states for calculations of background). Green curve is the intensity difference ($I_{obs} - I_{calc}$) between experimental data and calculations. Reduced $\chi^2 = 1.953$. The reflections are assigned to Re, Ne, and ζ-B; their *hkl* are designated in different colors for clarity.

**TABLE I**. Parameters of the equations of state of ζ-B compared to theoretical predictions. The EoSes are designated as follows: BM stays for the Birch-Murnaghan (3BM for the 3rd order, 2BM for the 2nd order), M for Murnaghan and V for Vinet.

| ζ-B | EoS | $V_{115}$, Å$^3$ | $K_{115}$, GPa | K´ | Ref. |
|---|---|---|---|---|---|
| **Experiment** | V | 39.19(8) | 577(65) | 4 | *present study* |
|  | 2BM | 39.20(8) | 575(66) | 4 | *present study* |
| *ab initio* calculations | 3BM | 38.99(4) | 626 | 3.6 | [19] *(Fan, 2014)* |
|  | 2BM | 39.11(4) | 674 | 4 | [18] *(Xu, 2015)* |
|  | M | 38.09(3) | 640 | 3.26 | [23] *(Ma, 2004)* |

# SUPPLEMENTARY MATERIALS.

**SUPPLEMENTARY TABLE I.** Some experimental details and single crystal X-ray diffraction and structural refinement data of γ-B and ζ-B determined in the present work.

| Parameter | γ-B | ζ-B |
|---|---|---|
| Pressure (GPa) | 58(1) | 115(2) |
| Empirical formula | $B_{28}$ | $B_8$ |
| Formula weight (g/mol) | 302.7 | 86.48 |
| Wavelength (Å) | 0.3738 | 0.29464 |
| Crystal system | Orthorhombic | Orthorhombic |
| Space group | *Pnnm* | *Cmce* |
| $a$ (Å) | 4.7521(16) | 2.7039(10) |
| $b$ (Å) | 5.295(2) | 4.8703(32) |
| $c$ (Å) | 6.4742(11) | 2.9697(6) |
| $V$ (Å$^3$) | 162.92(9) | 39.11(3) |
| Z | 1 | 1 |
| Calculated density (g/cm$^3$) | 3.08546 | 3.6715 |
| Crystal size (μm$^3$) | 5 × 6 × 8 | 8 × 7 × 10 |
| Theta range for data collection (deg) | 1.6543 to 17.0036 | 4.57 to 15.73 |
| Index range | -5 < h < 5 | -3 < h < 4 |
| | -5 < k < 6 | -6 < k < 5 |
| | -7 < l < 7 | -4 < l < 4 |
| Reflections collected | 388 | 88 |
| Independent reflections/$R_{int}$ | 177 / 0.0496 | 35 / 0.1499 |
| Refinement method | Least squares on $F^2$ | Least squares on $F^2$ |
| Data / restraints / parameters | 131 / 0 / 37 | 35 / 0 / 4 |
| Goodness of fit on $F^2$ | 1.056 | 1.293 |
| Final R indices [I>2σ(I)] | $R_1$ = 0.0676 | $R_1$ = 0.1368 |
| R indices (all data) | $R_1$ = 0.0854; $wR_2$ = 0.1645 | $R_1$ = 0.1783; $wR_2$ = 0.3106 |
| Largest diff. peak and hole (e/Å$^3$) | -0.52 and 0.43 | -0.50 and 0.55 |

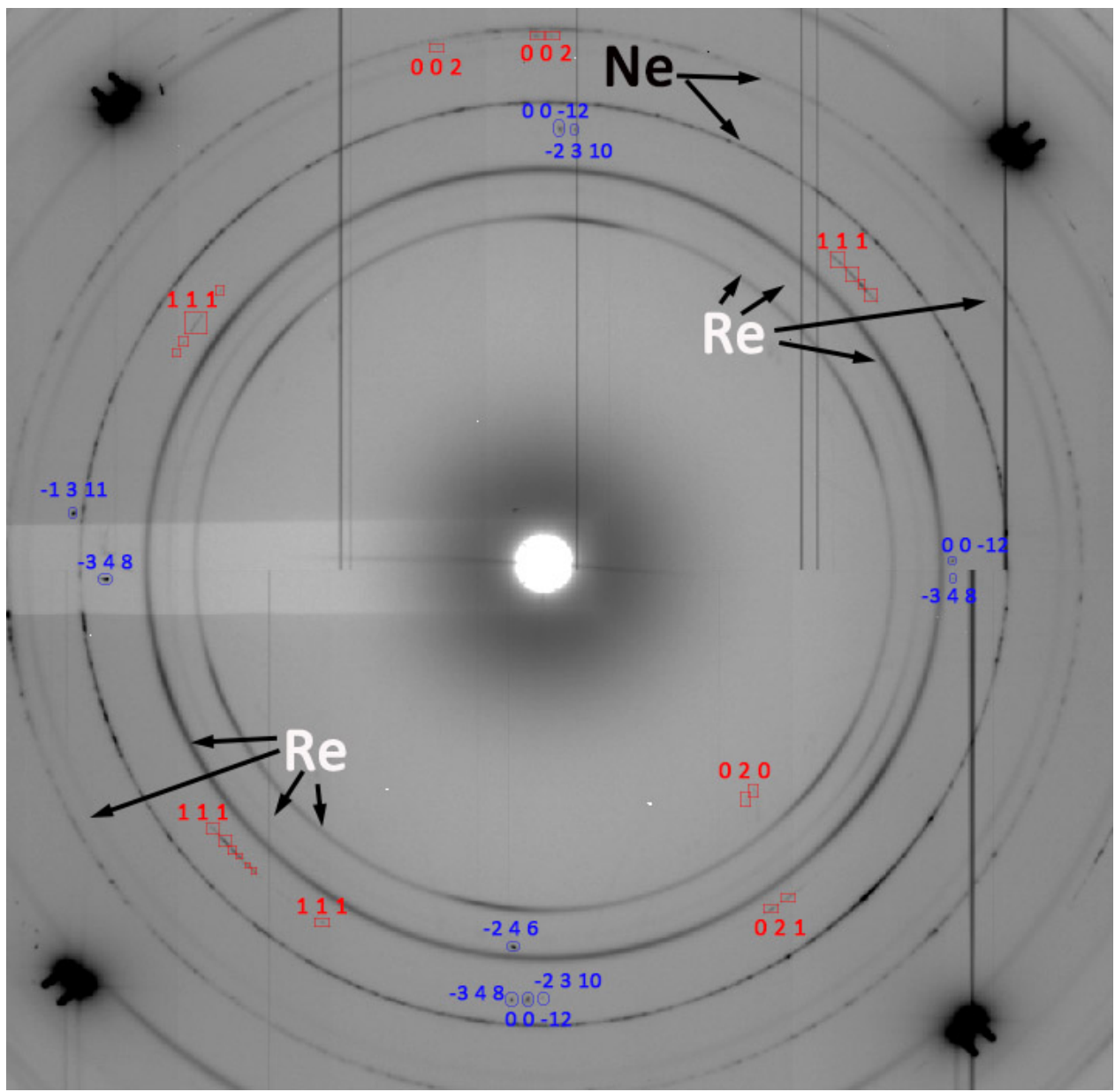

**SUPPLEMENTARY FIG. 1.** The 2D wide-scan X-ray diffraction image of ζ-B obtained at 115(2) GPa collected at P02.2 DESY and given as an example. Diffraction lines assigned to Re and Ne are pointed out by arrows. Blue circles mark the reflections related to untransformed β-B, red squares mark the reflections related to ζ-B. Black oversaturated reflections in the corners of the image are from the diamond anvils.

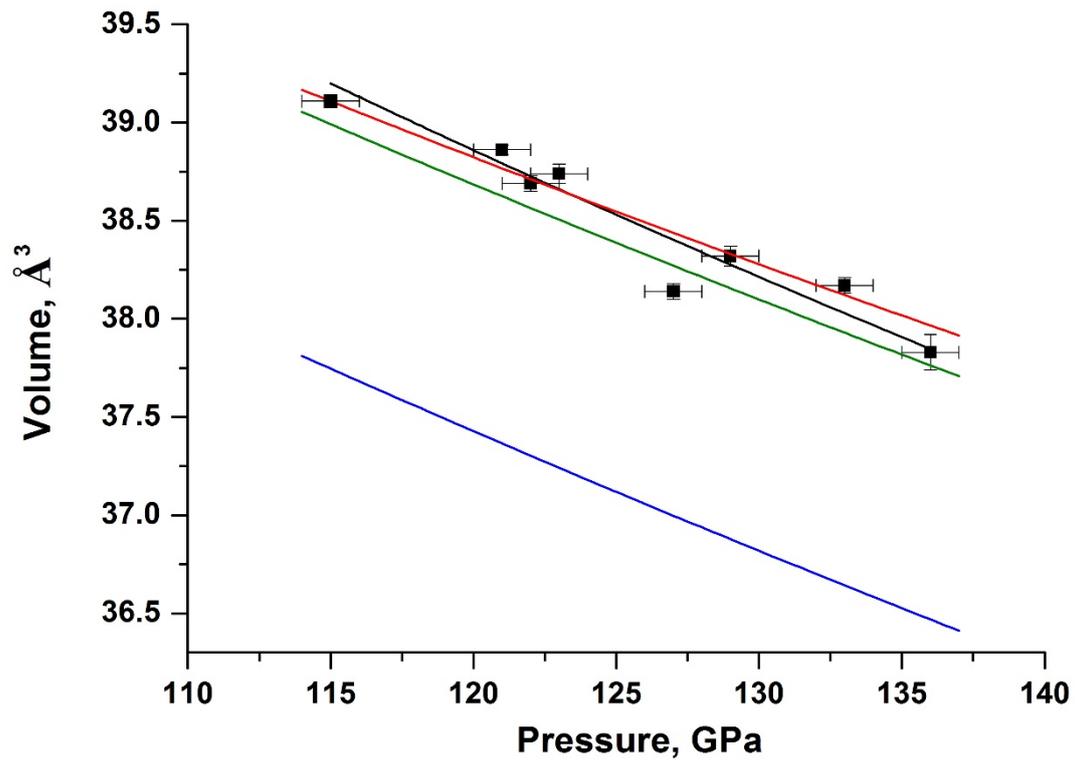

**SUPPLEMENTARY FIG 2.** The dependence of the unit cell volume of ζ-B on pressure. Black squares are our experimental points obtained from powder and single-crystal XRD data. Black line is the fit of our PV data with the 2BM EoS. Theoretically calculated EoSes are from the literature data [19] (green line, 3BM), [18] (red line, 3BM), and [23] (blue line, M).